\documentclass[sigconf]{acmart}

\usepackage[utf8]{inputenc}
\usepackage[T1]{fontenc}
\usepackage{geometry}
\usepackage{booktabs}
\usepackage{graphicx}
\usepackage{threeparttable}
\usepackage{array}
\usepackage{wasysym}
\usepackage{caption}
\usepackage[skip=1em]{caption}
\usepackage{graphicx,booktabs,threeparttable,amsmath}
\usepackage{graphicx}
\graphicspath{{assets/}}
\usepackage{framed,xcolor}

\providecommand{\LEFTcircle}{\symbol{"25D0}} 
\providecommand{\CIRCLE}{\symbol{"25CF}} 

\definecolor{shadecolor}{rgb}{0.9,0.9,0.9}

\usepackage{placeins}

\setcopyright{acmlicensed}
\acmYear{2026}
\copyrightyear{2026}
\acmDOI{XXXXXXX.XXXXXXX}

\acmConference[SIGCSE '26]{Proceedings of the 57th ACM Technical Symposium on Computer Science Education}{February 18--21, 2026}{St. Louis, MO, USA}
\acmBooktitle{Proceedings of the 57th ACM Technical Symposium on Computer Science Education (SIGCSE '26), February 18--21, 2026, St. Louis, MO, USA}

\newcommand{\tinysection}[1]{\noindent\textbf{#1:}\ }

\begin{document}

\title{CTF Archive:\protect\\ Capture, Curate, Learn Forever}

\author{Pratham Gupta}
\affiliation{
  \institution{Arizona State University}
  \city{Tempe}
  \state{AZ}
  \country{USA}
}
\email{prathamgupta@asu.edu}

\author{Aditya Gabani}
\affiliation{
  \institution{Arizona State University}
  \city{Tempe}
  \state{AZ}
  \country{USA}
}
\email{agabani@asu.edu}

\author{Connor Nelson}
\affiliation{
  \institution{Arizona State University}
  \city{Tempe}
  \state{AZ}
  \country{USA}
}
\email{Connor.D.Nelson@asu.edu}

\author{Yan Shoshitaishvili}
\affiliation{
  \institution{Arizona State University}
  \city{Tempe}
  \state{AZ}
  \country{USA}
}
\email{yans@asu.edu}

\renewcommand{\shortauthors}{Anonymous Author1, et al.}

\begin{abstract}
\small 
Capture the Flag (CTF) competitions represent a powerful experiential learning approach within cybersecurity education, blending diverse concepts into interactive challenges. However, the short duration (typically 24-48 hours) and ephemeral infrastructure of these events often impede sustained educational benefit. Learners face substantial barriers in revisiting unsolved challenges, primarily due to the cumbersome process of manually reconstructing and rehosting the challenges without comprehensive documentation or guidance. To address this critical gap, we introduce CTF Archive, a platform designed to preserve the educational value of CTF competitions by centralizing and archiving hundreds of challenges spanning over a decade in fully configured, ready-to-use environments. By removing the complexity of environment setup, CTF Archive allows learners to focus directly on conceptual understanding rather than technical troubleshooting. The availability of these preserved challenges encourages in-depth research and exploration at the learner's pace, significantly enhancing conceptual comprehension without the pressures of live competition. Additionally, public accessibility lowers entry barriers, promoting an inclusive educational experience. Overall, CTF Archive provides a scalable solution to integrate persistent, practical cybersecurity learning into academic curricula.
\end{abstract}



\maketitle

\section{Introduction}
The cybersecurity landscape changes rapidly with the discovery of new vulnerabilities and attack techniques which results in formation of new methods used to protect our current systems, yet formal courses and online materials often lag behind. As a result, learners struggle to master up-to-date topics or revisit older ones once forgotten which often creates a knowledge gap when the learners are solving real-world problems. Capture-the-Flag (CTF) competitions bridge this gap by packaging cutting-edge concepts which include topics like binary exploitation, cryptography, reverse engineering, web exploitation, osint, forensics, and miscellaneous. 
However, traditional CTFs run for only 24-48 hours, after which all challenges, their environments, and the embedded lessons disappear.

Capture-The-Flag exercises significantly increase student engagement and skill development in cybersecurity education~\cite{leune2017using}. This also creates a learning environment which incentivizes self-learning where learners do their own research while learning new concepts.
As Vykopal et al. discusses that integrating CTF games into coursework revealed key design issues around scoring, scaffolding, and analytics capabilities of platforms \cite{vykopal2020benefits}. We aim to help decrease the gap between the classical classroom and essential learning process through CTF challenges by preserving the knowledge.
With an emphasis on hands-on skills from cryptography to network security Švábenský et al. showed, there is often a neglect of human-aspect training \cite{svabensky2021cybersecurity}.

Even when past CTF challenges are made public usually in the form of GitHub repositories, rehosting them locally can be prohibitively time consuming.
Learners must wrestle with hardware mismatches, missing dependencies, unclear documentation, and Docker intricacies each step diverting attention away from the core objective of understanding the underlying security concept. Docker issues might demotivate the learner from understanding the concept due to problems that are not under their control. This setup overhead not only slows learning but often discourages deep, exploratory study across diverse topics.

In this paper we present the CTF Archive a centralized repository of 650+ CTF challenges spanning more than a decade, each pre-configured in browser-based Docker sandboxes. Learners interact via a familiar IDE and terminal, without manual installation or environment tweaking. By removing setup friction and preserving the full challenge context indefinitely, the CTF Archive empowers students to explore, practice, and retain cybersecurity concepts at their own pace. CTF Archive is also publicly available on GitHub for anyone to look at the challenge files and provide any feedback as well any problems that may arise in the form of issues. This encourages community collaboration and fosters interactive learning environment \cite{checkpoint2024}.
With the contributions page, it encourages the community to help in rehosting challenges which they think might be a good addition to the already available concepts.

\section{Motivation}
The project was started with an aim of being able to save the knowledge lost when the CTF Competitions concluded after 24-48 hours as is their nature. Once the competitions are over, the learners lose privilige to interact with the challenges. These challenges often contain "knowledge nuggets": concise, practical insights or problem solving techniques that teach participants something new. This knowledge preservation matters deeply in cybersecurity education, where hands-on exercises are the most effective way to internalize complex concepts. Even when these concept studded challenges are available to the public after the CTF competition has concluded, it still creates an intrinsic issue of rehosting those challenges with the technical hurdles inevitably slowing down the learning process where the focus is shifted from learning the concepts to understanding the docker images, hardware restrictions, and the tool dependencies.

\subsection{Accessibility}
By archiving hundreds of challenges of past CTF competitions in fully configured and tested dockerized containers mimicking the original challenge environment, our work aims to preserve these pedagogical assets indefinitely, eliminate setup friction, and empower learners to explore, practice, and retain critical cybersecurity skills. CTF Archive exists as a public repository on GitHub. CTF Archive not only safeguards valuable learning materials but also creates an accessible, scalable environment that aligns with evolving curricular needs—ensuring that every challenge remains a durable resource for self-directed education \cite{Erickson2017,Boettiger2015,Ahmed2019,Dooley2018,Vu2020}.

\subsection{Significant Work}
Creating these challenges with the hope of teaching new and different concepts is difficult as it requires a lot of knowledge but also difficulty assesment. This is identified in the retrospective of the DEF CON Cloud Village CTF, organizers "had to do a ton of work creating challenges nearly from scratch" \cite{chickowski2024}.
The same idea is also approached by Tao Sauvage, creating Hammercon 2025's CTF was “a labor of love” involving 16 original challenges, beta-testing, and solo design effort \cite{sauvage2025}.

CTF organizers spend hundreds of hours spanning over months of work to create these challenges, but in most cases they are taken down once the CTF is offline. By archiving them we are also preserving and acknowledging the work done by challenge authors.   

\subsection{Preserving Knowledge}
As CTF competitions take months of preparation to make sure knowledge nuggets exist, which they learn while solving these challenges.
The preservation of these challenges gives anyone an opportunity to learn from people who spend hours building a challenge. Our platform caters to their needs while making sure the process of learning is authentic and fun.

This journey of self-exploration would give anyone the freedom to choose and have a feeling of participating in any past event as the event would still exist in an archived form in CTF Archive.

\subsection{Barrier to Entry}
The time spent in rehosting a challenge on your local environment which is seen by some as an opportunity to learn rehosting, is often a challenge to students who want to learn a topic rather than delve into the intricacies of challenge development and rehosting.

During live CTFs the challenges are ready to be worked on by the participants. We want to replicate that same environment bridging the gap where any learner who wants to learn has the challenge running and ready to be solved. They also get the freedom to choose the concepts they want to learn without any rehosting concerns.

\section{Related Work}
\label{sec:related}
Several projects tackle the problem of preserving live CTF challenges, either by archiving challenge source files or by keeping the CTF website accessible after the event ends. This section discusses and compares these platforms with ours.

\textbf{Sajjadium's CTF archive repository \cite{sajjadium2025ctfarchives}, r3kapig's Notion\cite{r3kapig2025indexdocsformat}, and Shellstorm \cite{shellstorm2025repoCTF}} are all notable resources that archive and provides access to past CTF challenges. Sajjadium's archive is publicly hosted on GitHub, providing an extensive collection of source files from multiple CTFs spanning many years. r3kapig's archive uses notion to systematically maintain challenges from multiple CTFs organized in the past 3 years. Shellstorm is another prominent project that archives various CTF challenges, with some of the CTFs dating more than a decade ago. Despite their utility, all three platform faces similar challenges. Primarily, they lack proper documentation on setup of these challenge which adds another difficulty layer of rehosting them. Moreover, some of the challenges are missing dependency files which makes it difficult to track what's missing or why the challenges are not working. Sometimes the challenge source files are broken which can be due to missing libc files--core libraries from Linux GNU necessary to build and run any file or even kernel incompatibility which cannot be discovered easily.

\textbf{Ångstrom CTF \cite{angstromctf2025}, Imaginary CTF \cite{imaginaryctf2025}, OOO's Archive \cite{archiveooo2025}, and Cryptohack \cite{cryptohack}:} Ångstrom CTF provides an archive of challenges hosted by the organizing team offering challenge source files and descriptions. Imaginary CTF is an archive hosted by the ImaginaryCTF organizing team, providing challenge source files, descriptions, flags, and write-ups. OOO's archive is a playable archive of DEFCON CTF challenges organized organized by the Order of the Overflow from 2018-2021, containing all challenge source files on either live servers or with clear instructions for local hosting. Cryptohack is a platform featuring playable cryptography challenges organized in categories, complete with source files and descriptions. Although all of these platforms offer organized archives and most allow flag submissions(ImaginaryCTF being the exception here), none of them provide an integrated environment for solving the challenges. Additionally, these archives primarily rehost their own past CTF challenges. 

The closest equivalent platform to CTF Archive is \textbf{PicoGym}, which offers past challenges organized by the \textbf{PicoCTF} team. It offers a web-based shell to solve certain challenges, as well as a persistent storage directory with live servers and flag submission. Although this platform provides tooling and environment for solving challenges, not all challenges are solvable with their web-based shell, and migrating the challenge files from their website to their web-based shell requires basic linux command line knowledge. CTF Archive however provides a platform where all the challenges are playable while having extensive tools with persistent storage.\cite{chapman2014picoctf}

\textbf{CTFtime \cite{ctftime}} is an database of CTF events complete with list of events complete with all the metadata providing team rankings, CTF rankings based on difficulty which is widely by the cybersecurity community as a resource. It also features a platform where individuals can upload their writeups CTFs after the CTFs are concluded. While this platform is an exceptional resource, it doesn't necessarily archive all challenge source files and rehost instructions. Since the writeups are community based they are often scant. CTF Archive has rehost instructions and challenge source files making it different than ctftime.

\textbf{CTFd \cite{ctfd}} is a popular open-source framework for running live CTF events.  Organizers can upload challenges, manage scoring, and host a dynamic leaderboard within a Dockerized instance. While CTFd excels at powering time-boxed competitions and offers plugins for custom scoring or hints, it inherently assumes a live event context: once the contest ends, many deployments are taken offline or archived without an accessible interface. Moreover, participants typically must install local toolchains or SSH into remote VMs to solve problems, reintroducing setup complexity and dependency issues that hinder self-paced learning. As CTF Archive is hosted on pwn.college, it offers us all the funtionality offered by CTFd due to CTFd being used to develop pwn.college.

\textbf{GitHub repositories:} Numerous CTF organizing teams make their challenges public after they are concluded via Github repositories sometimes including writeups. These often come with proper instructions on rehosting the them. While the source files and documentation exist, they do not provide an IDE to solve the challenges. Additionally, some of the challenges are hosted on Dockerhub when not maintained making the rehosting process difficult as users have to spend time understanding the context often resulting in the challenges being not rehostable. CTF Archive removes the rehosting hurdle while providing an IDE.

\textbf{Professional Projects:} There are commercial platforms available that have their own challenges avaiable for users to learn through. Platforms like HacktheBox \cite{hackthebox}, TryHackMe \cite{tryhackme}, and RET2 Wargames \cite{ret2wargames} also provide a working environment for users to work on, but the major hurdle with these platforms is not all the features are available with the free tier account pushing a paid resource. Furthermore, these platforms do not archive other CTF challenges, their main purpose is to teach and train cybersecurity in a systematic manner whereas the main goal of CTF archive is to preserve the knowledge taught by CTF challenges.

\textbf{CTF Archive's novelty:} These archives highlight the importance of preserving CTF challenges while revealing a gap in their rehosting. This is where CTF Archive plays a crucial role, providing a pre-configured environment, with the necessary tools and an integrated interface for solving the challenges. 

We help bridge the aforementioned gap by archiving various different CTFs spanning hundreds of challenges. This helps in the learning process by eliminating the barrier of manually rehosting.
Another difference between CTF Archive and all other platforms is that all the challenges are dockerized with the intention of maintaining them indefinitely.

\begin{table}[!htbp]
  \centering
  \footnotesize
  \label{tab:features}
  
  \newcommand*\rot[1]{\hbox to1em{\rotatebox[origin=bl]{60}{#1}}\hss}
  \newcommand*\feature[1]{\ifcase#1 -\or\LEFTcircle\or\CIRCLE\fi}
  
  \newcommand*\fa[1]{\feature#1}
  \newcommand*\fb[2]{\feature#1 & \feature#2}
  \newcommand*\fc[3]{\feature#1 & \feature#2 & \feature#3}
  \newcommand*\fd[4]{\feature#1 & \feature#2 & \feature#3 & \feature#4}
  \newcommand*\fg[7]{\feature#1 & \feature#2 & \feature#3 & \feature#4 & \feature#5 & \feature#6 & \feature#7}
  \makeatletter
  \newcommand*\ex[5]{#1 & \fd#2 & \fd#3 & \fc#4 & \fg#5 \\}
  \makeatother

  \resizebox{0.9\columnwidth}{!}{%
    \begin{threeparttable}
      \begin{tabular}{ 
            l 
        !{\kern0em} c@{}c@{}c@{}c 
        !{\kern0em} c@{}c@{}c@{}c 
        !{\kern1.5em} c@{}c@{}c 
        !{\kern2em} c@{}c@{}c@{}c@{}c@{}c@{}c 
      }
        \toprule
        \textbf{Project}
        & \multicolumn{4}{l}{\textbf{Source}}
        & \multicolumn{4}{l}{\textbf{Openness}}
        & \multicolumn{3}{l}{\textbf{Interaction}}
        & \multicolumn{7}{l}{\textbf{Environment}} \\
        \midrule
        & \rot{Challenge Files} 
        & \rot{Rehost Instructions} 
        & \rot{Writeups} 
        & \rot{Other CTF Challenges} 
        & \rot{Open Source}
        & \rot{Deployable}
        & \rot{Free To Use}
        & \rot{Flag Submission}
        & \rot{Netcat}
        & \rot{Website}
        & \rot{In-Environment}
        & \rot{SSH}
        & \rot{In-Browser Terminal}
        & \rot{In-Browser Code Editor}
        & \rot{In-Browser Desktop}
        & \rot{Persistent Data}
        & \rot{Rich Tooling}
        & \rot{Privileged Access} \\
        \midrule
        \ex{Sajjadium's GutHub archive}    {2 1 1 2} {2 0 2 0} {0 0 0} {0 0 0 0 0 0 0}
        \ex{r3kapig's Notion archive}      {2 1 1 2} {2 0 2 0} {0 0 0} {0 0 0 0 0 0 0}
        \ex{ShellStorm}                    {2 0 0 2} {2 0 2 0} {0 0 0} {0 0 0 0 0 0 0}
        \midrule
        \ex{Ångstrom CTF}                  {2 2 0 0} {0 1 2 2} {1 2 0} {0 0 0 0 0 0 0}
        \ex{ImaginaryCTF}                  {2 2 2 0} {0 1 2 0} {1 2 0} {0 0 0 0 0 0 0}
        \ex{archive.ooo}                   {2 2 0 0} {2 2 2 2} {1 2 0} {0 0 0 0 0 0 0}
        \ex{Cryptohack}                    {2 2 0 0} {0 2 2 2} {2 0 0} {0 0 0 0 0 0 0}
        \ex{PicoCTF}                       {2 2 0 0} {0 2 2 2} {2 2 1} {0 2 0 0 2 2 0}
        \midrule
        \ex{Public GitHub repos}           {2 1 1 2} {2 0 2 0} {0 0 0} {0 0 0 0 0 0 0}
        \ex{CTFtime}                       {1 1 1 2} {0 0 2 0} {0 2 0} {0 0 0 0 0 0 0}
        \ex{CTFd}                          {0 0 0 0} {2 1 2 0} {1 2 1} {0 0 0 0 0 0 0}
        \midrule
        \ex{\texttt{CTF Archive}}          {2 2 0 2} {2 2 2 2} {1 2 2} {2 2 2 2 2 2 2}
        \bottomrule
      \end{tabular}

      \begin{tablenotes}
        \item[\CIRCLE] provides property; 
        \item[\LEFTcircle] partially provides property; 
        \item[-] does not provide property.
      \end{tablenotes}
    \end{threeparttable}%
  }
  \vspace{0.3em}
  \caption{This table shows the different features and functionalities of various projects when compared to CTF Archive. ~\ref{sec:related}.}
  \vspace{-3.5em}
\end{table}

\FloatBarrier

\section{CTF Archive Framework}
Our design centers on three goals: preserve authors’ pedagogical intent, remove setup friction, and support authentic, self-paced practice. We deploy on a browser-based teaching platform that provides a VS Code interface, terminal, optional GUI desktop, and persistent storage~\cite{nelson2024dojo}.

\begin{figure}[ht]
  \centering
  \includegraphics[width=\linewidth]{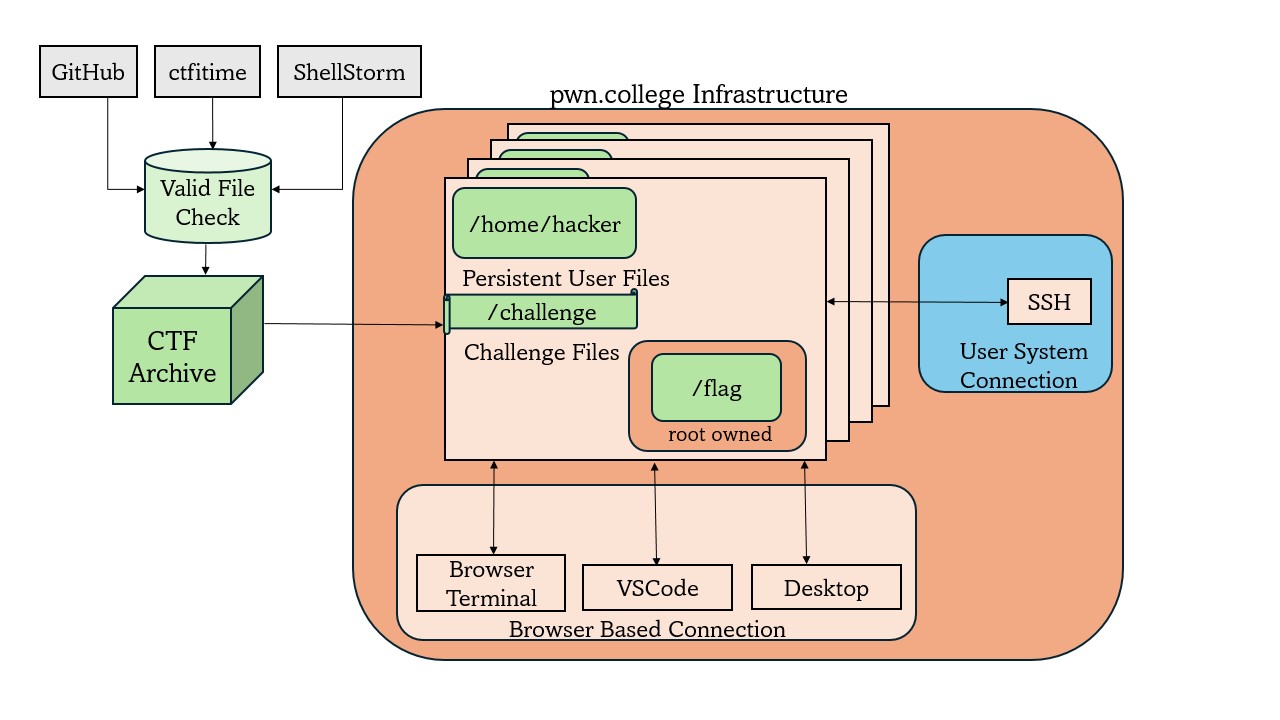}
  \caption[CTF Archive Infrastructure]{\textbf{CTF Archive Infrastructure.} This diagram shows the sequence from challenge design through deployment.}
  \label{fig:ctf-workflow}
  \vspace{-2em}
\end{figure}

As shown in Figure~\ref{fig:ctf-workflow}, our pipeline preserves upstream behavior by reconstructing original environments and documenting deviations in \texttt{REHOST.md}.

\subsection{Infrastructure}
pwn.college is public platform maintained by a team at ASU, we chose it due to its ability to host community dojos. 
pwn.college uses docker containers to host challenges which was a huge advantage as docker provides a sandbox or essentially a closed environment to run a challenge mirroring the original challenge environment from the CTF competitions.\cite{nelson2024dojo}
The ability to host our own dojo with a compilation of different CTFs with defined modules containing the challenges.
pwn.college's environment and unconstrained challenge building structure gave us the ability to rehost the challenges as close to the original environment as possible.
The platform also provides a community discord server -- a social messaging platform, containing over 22,000+ users with more than 40,000 users on the platform itself, with an ability to create your own forum for community dojos. This acts as a communication method between us and our users, for them to let us know any challenges issues or fixes while providing them an option to help us in rehosting them through GitHub pull requests. This also contributes towards community engagement, users can work together and tackle the problems, users struggling can gain help from the community. Another feature of the platform is its ability to provide us user solve data which shows solve information which uses profile user ids.  
Karagiannis and Magkos show that adapting CTF challenges into virtual cybersecurity learning environments significantly enhances student engagement and skill acquisition \cite{karagiannis2020adapting}.

\subsection{Architecture and Workflow}
For each challenge we: (1) collect context (event, year, category, nominal difficulty/points); (2) reconstruct the environment, preferring original Dockerfiles and inferring dependencies as needed; (3) harden and document with a per-challenge \texttt{REHOST.md}; and (4) package and deploy as containerized services with uniform entry points (terminal, netcat/HTTP endpoints) and persistent storage. To avoid a single bloated image, we use a small base with common tooling and thin per-challenge layers via multi-stage builds.

\subsection{Uniform Tooling and Interfaces}
All challenges expose an in-browser terminal and editor; optional GUI desktop; network endpoints where appropriate; and persistent storage. Where upstream flag checkers are unavailable and flags are embedded, \emph{FlagCheck} verifies user submissions against a stored hash to preserve solving intent without revealing answers.

\subsection{Approach}
\tinysection{Challenge Collection \& Context} We went through ctftime for metadata (competition name, year, category, point value) on numerous past CTFs. This contextual information helped in labelling the challenges in terms of difficulty and helped learners set expectations.

\tinysection{Preserving Original Environments}For each challenge, we tried to locate official Dockerfiles or source assets in public GitHub repositories and online individual platforms. Where documentation existed, we rehosted the exact Docker image; where it was missing, we reverse-engineered dependency lists and consulted community write-ups to reconstruct the intended toolchain.

\tinysection{Unified Tooling \& Sandbox} All challenges run in a single Docker image maintained by CTF Archive. We augmented pwn.college's base image with any obscure or challenge-specific tools (e.g., custom compilers, debuggers, crypto libraries), ensuring that learners never encounter “missing dependency” errors.

\tinysection{Difficulty \& Scoring} We preserved original point values assigned by challenge authors and surfaced these scores in the pwn.college interface. This allows learners to choose problems aligned with their skill level and expected time investment.

\tinysection{FlagCheck}During the Rehosting process of certain CTF challenges we were presented with a unique problem -- for some of the challenges the only available challenge source files were executable binaries and the flag is embedded in the binaries. For uniformity and for preserving the true value of the challenge, we decide to use our executable binary called FlagCheck which uses the challenge flag's SHA256 hash to verify if the user provided flag is correct.

\begin{figure}[ht]
  \centering
  \includegraphics[width=\linewidth]{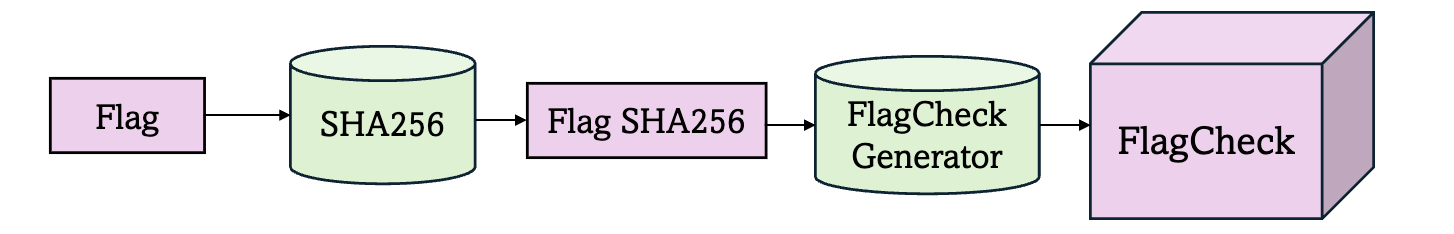}
  \caption[CTF Archive Custom FlagCheck]{\textbf{CTF Archive Custom FlagCheck-} This is the process used to create flagChecks, first the challenge's actual flag is hashed using sha256, then it is put inside the flagCheck generator which compares the actual flag with the user input to provide the pwn.college flag.}
  \label{fig:FlagCheck-generation}
\end{figure}

\tinysection{Comprehensive Rehosting Documentation}Each challenge includes a \texttt{REHOST.md} file detailing step-by-step instructions—from cloning source files to building and verifying the Docker image—so learners and instructors can reproduce or adapt our work without any issues.

\begin{table}[t]
  \centering
  \small
  \caption{Coverage and solves by challenge type.}
  \label{tab:challenge-coverage}
  \begin{tabular}{lrr}
    \toprule
    \textbf{Challenge Type} & \textbf{Available} & \textbf{Solves} \\
    \midrule
    Cryptography & 261 & 5{,}204 \\
    Binary Exploitation (PWN) & 168 & 504 \\
    Reverse Engineering & 125 & 856 \\
    Web Exploitation & 26 & 220 \\
    Forensics & 43 & 160 \\
    OSINT & 17 & 71 \\
    Blockchain & 2 & 2 \\
    Radio Frequency & 2 & 8 \\
    Social Engineering & 1 & 2 \\
    Steganography & 1 & 4 \\
    MISC & 54 & 364 \\
    \midrule
    \textbf{Total} & \textbf{700} & \textbf{7{,}395} \\
    \bottomrule
  \end{tabular}
\end{table}


\noindent
\textbf{Deployment Context and Metrics:}  
\noindent
This transparent, repeatable approach demonstrates that the CTF Archive's design aligns directly with our educational goals: preserving challenge intent, lowering technical barriers, and embedding continuous formative assessment into the learner experience. The amount of CTFs proportionally increase the amount of different challenges resulting in an increase in the attraction gained from the community.

\section{Impact}
This section covers the impact of CTF Archive on key domains like community, academia, and research. We discuss the community support for an open source project which provides all the necessary documentation to contribute. Next, we show how integrating our challenges into a large-enrollment cybersecurity course has demonstrably enhanced student engagement and learning outcomes. We then quantify the surge in student uptake—particularly in cryptography exercises and examine how that growth has driven further platform enhancements. Finally, we explore emerging research uses of our comprehensive, Docker-ready dataset, from large-language-model training to automated solvability verification. Together, these contributions show how CTF Archive not only lowers the barrier to hands-on cybersecurity education but also contributes to new research and community participation.

\subsection{Community}
The CTF archive is an open source project on Github, contributions from the community are welcomed. We provide documentation to anyone who wants to contribute to the project, which helps us maintain our structure and the users contribute without much of a hassle. Till date we have had 110+ issues and PRs posted, out of which 70+ issues have been resolved, helping us maintain and create a better resource for the community. An additional benefit of being an open source project is our documentation for individual challenges allowing users to manually rehost them if needed.

\subsection{Academia}
CTF Archive was used as a part of a required cybersecurity course in a large public R1 university amassing a thousand students during the Fall 2024 semester into using the challenges rehosted as a way of gaining extra credit as well as reinforcing the learned concepts using new methods which might not be incorporated in the curriculum.\cite{morgan2021integrating}

\begin{figure}[h]
  \centering
  \includegraphics[width=\linewidth]{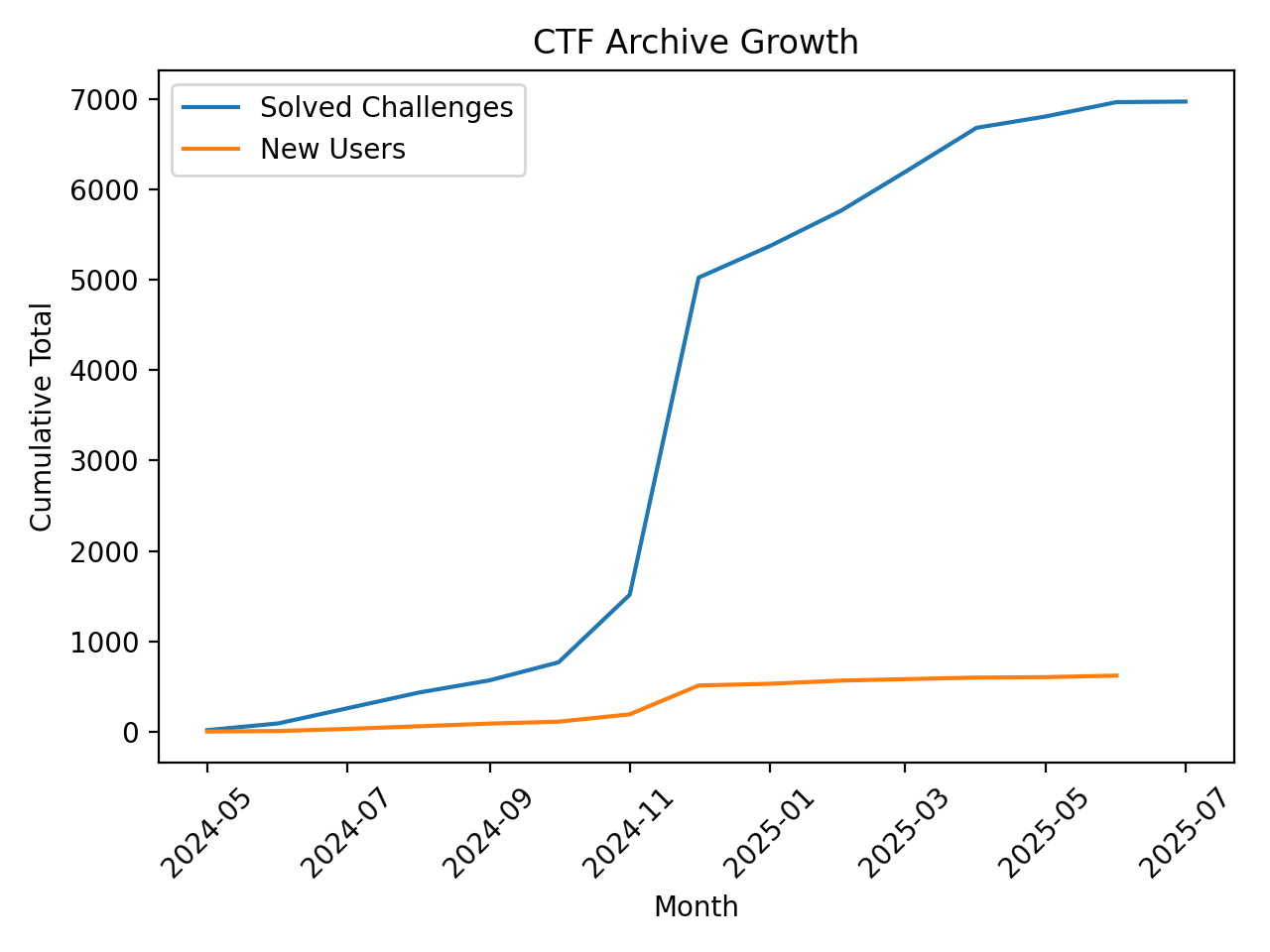}
  \caption[CTF Archive Combined Growth Graph]{
  
  \textbf{CTF Archive Combined Growth-} This graph shows the new user growth as well as the solves. As its visible that during the December 2024 timeframe, the amount of solves to user ratio is higher than what before. This is the direct result of CSE 365 students joining and solving challenges.}
  \label{fig:solves-growth}
\end{figure}

\subsubsection{\textbf{Students:}}
This resulted in an exponential growth in solving challenges during the month of December which correlates to the end of Fall semester. The majority of the solves were for cryptography challenges as they were by design the first challenges they would encounter on the page.\cite{sanders2019ctf}
As we saw this increase in recognition of CTF Archive, we also got requests to add more CTFs, with some community members helping in rehosting CTFs, and the resolved caused inconveniences in the challenges due to reasons like not having flagChecks, platform infrastructure changes, lack of initial comprehensive testing which was later corrected by adding manual checks.

\subsection{Research}
CTF Archive has already begun to catalyze a diverse set of investigations using our large dataset with interest from different teams aiming to incorporate CTF Archive into their project.

\subsubsection{Potential Research} A potential use case of CTF Archive we envision is the dataset, with currently having 650+ challenges and plans on adding more. Currently, there are multiple research studies ongoing using Large Language Models(LLM) to find systemd vulnerabilities and different exploitation techniques. With the availability of our dataset which includes challenges teaching multiple concepts while requiring different kinds of exploitation techniques, it can be used to train these LLM models potentially helping find system vulnerabilities and exploitation techinques efficiently.

\subsubsection{Research Interest} There was also interest from a researcher from a large public university member of GO8 regarding the implementation of CTF Archive in a dockerized headless environment where our manual and automatic checks with the help of flagChecks was able to provide challenge solvability verification. This provided an important use case where CTF Archive can be used as a tool for various research projects for any researcher.

\subsubsection{\textbf{Research Team Collaboration:}} The \textbf{CAISI Cyber Evaluations framework} developed by the \textbf{National Institute of Standards and Technology (NIST)} incorporated CTF Archive as an externally-tracked benchmark using \textbf{UK AI Security Institute's AISI framework} for large language model evaluations using our 650+ challenges rehosted from real world CTF competitions for the agents being used by them. This is one of the use cases mentioned earlier in the potential research use cases for CTF Archive as it provides a comprehensive list of challenges that can be used with minimal effort due to the documentation and guides provided.\cite{caisicyberevals}\cite{ukaisecurityinstitute}

\subsection{Reflection}
As we reflect on the challenges and learning experience gained by the learners we understand the need of such a collection is absolutely paramount to preserving the value of the knowledge CTFs hold. This increases the opportunity for students who are just starting their learning journey, students whose interest motivates them to learn something out of their current field, and help seasoned Cybersecurity professionals polish their skills.
This is not just a collection of challenges but its a library of potential tangents to different learning paths giving anyone who wants to learn a chance to follow what excites them as well as giving them a new way to look at certain problems paving a path to new discoveries in critical cybersecurity areas.

\section{Limitations}
Rehosting is two edged sword where doing it manually helps in gaining insights about how challenges work while learning new concepts and techinques, but it also hinders the learning process for the users who do not want to spend the time they wanted to learn a new concept on rehosting. For example, for a user not familiar with docker, rehosting a challenge that is hosted via a docker helps the user learn an entirely new concept. Similarly, fixing errors, writing new scripts to replicate the challenge environment, teaches concepts that the challenge author never intended. Since we provide all the documentation to rehost, CTF Archive gives user an option to either use the rehosted challenges or do it themselves.

\section{Conclusion}
In this paper we have demonstrated how the CTF Archive preserves and amplifies the pedagogical value of Capture-the-Flag competitions. By rehosting over 650 real-world challenges in fully preconfigured, browser-based Docker sandboxes, CTF Archive overcomes two major barriers to deep learning: the ephemeral nature of live CTFs and the technical friction of manual environment setup. 

We made sure our chosen platforms features align closely with established best practices for active learning and skills development, as documented in prior studies of CTF-based education. \cite{trickel2017shell}. Learners reported in the survey that the preconfigured environments removed setup obstacles, increased their confidence in tackling unfamiliar vulnerabilities, and enabled iterative review of key concepts long after original contests concluded.

\end{document}